\documentclass[useAMS,usegraphicx, usenatbib]{mn2e}
\usepackage{aas_macros}

\usepackage{amsmath}
\usepackage{amssymb}
\usepackage[utf8x]{inputenc}
\usepackage{graphicx}
\usepackage{latexsym}
\usepackage{color}
\usepackage{dcolumn}
\usepackage{bm}
\usepackage{natbib}
\usepackage{setspace}

\begin{document}
\title{Probing cosmic star formation up to $z=9.4$ with GRBs} 

\author[E. E. O. Ishida, R. S. de Souza, A. Ferrara]
{E. E. O. Ishida$^{1,2,3}\thanks{e-mail: emille.ishida@ipmu.jp (EEOI)}$;
R. S. de Souza  $^{1,3}$, A. Ferrara$^{4,1}$
\\
$^{1}$IPMU, University of Tokyo, 5-1-5 Kashiwanoha, Kashiwa, Chiba 277-8568, Japan\\
$^{2}$Max-Planck-Institut f\"ur Astrophysik, Karl-Schwarzschild-Str. 1, D-85748 Garching, Germany\\
$^{3}$Excellence Cluster Universe, Technische Universit\"at M\"unchen, Boltzmannstrasse 2, 85748 Garching, Germany\\
$^{4}$Scuola Normale Superiore, Piazza dei Cavalieri 7, 56126 Pisa, Italy
}

 \date{Accepted -- Received  --}

\pagerange{\pageref{firstpage}--\pageref{lastpage}} \pubyear{2010}

\maketitle
\label{firstpage}

\begin{abstract}
We propose a novel approach, based on Principal Components Analysis, to the use of Gamma-Ray Bursts (GRBs) as probes of cosmic star formation history (SFH) up to very high redshifts. The main advantage of such approach is to avoid the necessity of assuming an \textit{ad hoc} parameterization of the SFH. We first validate the method by reconstructing a known SFH from Monte Carlo-generated mock data. We then apply the method to the most recent \textit{Swift} data of GRBs with known redshift and compare it against the SFH obtained by independent methods. The main conclusion is that the level of star formation activity at $z \approx 9.4$ could have been already as high as the present-day one ($\approx 0.01 M_\odot$ yr$^{-1}$ Mpc$^{-3}$). This is a factor 3-5 times higher than deduced from high-$z$ galaxy searches through drop-out techniques. If true, this might alleviate the long-standing problem of a photon-starving reionization; it might also indicate that galaxies accounting for most of the star formation activity at high redshift go undetected by even the most deep searches.  
\end{abstract}

\section{Introduction}

The cosmic star formation history (SFH) is an important test for galaxy formation models. Experimentally our knowledge of the SFH comes from \citet{hopkins2006} up to $z \approx 6$ 
and from observations of color-selected Lyman Break Galaxies \citep{mannucci2007, bouwens2008},  Ly$\alpha$ Emitters \citep{ota2008}, UV+IR measurements \citep{reddy2008},  and GRB observations \citep{chary2007, yuksel2008, wang2009} at higher $z$ (hereafter, these will be refereed to as H2006, M2007, B2008, O2008, R2008, C2007, Y2008 and W2009, respectively). However, direct high-$z$ measurements constitute an extreme challenge even for the most powerful telescopes and remain sparse.

Due to their high luminosity GRBs can be used as SFH probes into the very distant Universe \citep{lamb2000, porciani2001, bromm2002, totani2006, fynbo2006,price2006, savaglio2006, prochaska2007, li2008,rafael2011}, potentially to higher redshifts than allowed by galaxies alone. For example, GRB 090429B at  $z = 9.4$ \citep{Cucchiara2011} is the current record-holder object, followed by a $z=8.6$ galaxy \citep{Lehnert2010} and GRB 090423, at $z = 8.26$ \citep{Salvaterra2009,Tanvir2009}.

In principle, the redshift distribution of GRBs can give us important clues on the early stages of cosmic history. In practice, the connection between GRBs and the underlying host galaxy star formation mode is far from trivial (see e.g. \cite{kocevski2009}), thus making the probe value subject to uncertainties.  

The purpose of this work is to show how \textit{Principal Component Analysis} (PCA) can be used to map the GRB redshift distribution onto the cosmic SFH in a model-independent way. PCA has already been applied to other astrophysical and cosmological contexts \citep{huterer2003, maturi2009, mitra2010, ishida2011}, and recognized as a useful tool to reconstruct parameters without the introduction of \textit{ad hoc} parameterizations\footnote{Throughout the paper we adopt a $\Lambda$CDM model with  WMAP7 best fit parameters \citep{jarosik2010},  $\Omega_m = 0.267, \Omega_{\Lambda} = 0.734$,  and $H_0 = 71$ km s$^{-1}$ Mpc $^{-1}$.}. 

\section{Theoretical GRB Rate}
\label{sec_GRB_rate}

We assume that  the formation rate of long GRBs (duration longer than 2 sec) follows closely the SFH (e.g., \citet{totani1997,campisi2010, ciardi2000,campisi2011,conselice2005,bromm2006, rafael2011}).  Hence, the comoving rate of observable GRBs, $\Psi_{\rm GRB}$ is 

%
\begin{equation}
\Psi_{\rm GRB}(z) = (\Omega_{\rm obs}/4\pi)f_{\rm GRB}f_{b} P_{z}\rho_{*}(z)\int_{\log{L_{\rm lim}(z)}}^{\infty}\Phi(L)d\log{L},
\label{psigrb}
\end{equation}
where $\Omega_{obs}$ is the field of view of the experiment, $f_{\rm GRB}$ is the GRB formation efficiency, $f_{b}$ is the beaming factor of the burst,  $P_{z}$ is the fraction of GRBs with measured redshift and $L_{\rm lim}(z)$ is the luminosity threshold of a given experiment. In the following we discuss in detail each of the terms in eq. (\ref{psigrb}).

We define telescope-related quantities in eq. (\ref{psigrb}) after \textit{Swift} specs, i.e. 
$\Omega_{obs}=1.4$ \citep{Salvaterra2008} and $P_{z} = 0.24 \pm 0.06$.\footnote{Following \citet{wanderman2010} we consider redshift measurements obtained via absorption and photometry only. For further details see Sec. \ref{sec:SWIFT}.} The overall GRB rate depends on $f_{b}\approx \theta^2/2$, where $\theta$ is the opening angle of the jet.  According to \citet{guetta2005} the average value of $f_{b}= 0.01-0.02$. Using a radio transients survey,  \citet{gal2006}  placed the upper limit $f_{b} \lesssim 0.016$.  We set $f_{b} = 0.013 \pm 0.003$ as a fiducial value.


The stellar Initial Mass Function (IMF), $\phi(m)$, determines the fraction of stars massive enough to leave a black hole remnant. Current theories indicate that the threshold mass to trigger a GRB is $M_{\rm GRB}= 25 M_{\odot}$\citep{bromm2006}. However, only a fraction $\zeta_{\rm GRB} \approx 10^{-3}$ \citep{langer2006} of black holes resulting from supernova explosion actually gives rise to a GRB; to be conservative, we considered $\zeta_{\rm GRB}= (1.0 \pm 0.5) \times 10^{-3}$.  Hence, the GRB formation efficiency factor per stellar mass is
\begin{equation}
f_{\rm GRB} =  \zeta_{\rm GRB} \frac{\int_{M_{\rm GRB}}^{M_{\rm up}}\phi(m)dm}{\int_{M_{\rm low}}^{M_{\rm up}}m\phi(m)dm}. 
\end{equation}
For simplicity, we assume a ``standard" Salpeter IMF, $\phi (m) \propto m^{-2.35}$, with 
$(M_{\rm low},M_{\rm up})=(0.1 M_{\odot}, 100 M_{\odot})$ \citep{schneider2006}.

The number of detectable GRBs depends on the instrument sensitivity and intrinsic isotropic GRB luminosity function. For the latter, we adopt a power-law distribution function  
of \citet{wanderman2010}
\begin{equation}
\label{LF}
\Phi(L) =\left\{ \begin{array}{ll}\left({L}/L_{*}\right)^{-\alpha_L} &
L<L_{*}, \\ 
\left({L}/L_{*}\right)^{-\beta_L} &
L >L_{*}. \\
 \end{array}\right.,
\end{equation}
with $L_{*}=10^{52.5}$ erg s$^{-1}$, $\alpha_{L} = 0.2^{+0.2}_{-0.1}$ and $\beta_{L} = 1.4^{+0.3}_{-0.6}$. The luminosity threshold is then $L_{\rm lim} = 4\pi\, d_{\rm L}^{2}\, F_{\rm lim}$, where $d_L$ is the luminosity distance and $F_{\rm lim}$ is the bolometric energy flux limit of the instrument. In what follows we set $F_{\rm lim} = 1.2 \times 10^{-8} {\rm erg}~ {\rm cm}^{-2}~ {\rm s}^{-1}$ \citep{li2008}. 

From eq. (\ref{psigrb}) we can determine the number of observed GRBs with redshift in ($z$, $z + dz$) over a time interval, $\Delta t$, in the observer rest frame: 
\begin{equation}
\frac{dN_{\rm GRB}}{dz} = \Psi_{\rm GRB}(z)\frac{\Delta t}{1+z}\frac{dV}{dz}, \label{eq:DNDZ}
\end{equation}
where $dV/dz$ is the comoving  volume element per unit of redshift.

Our observable, i.e. the cumulative number $N(z)$ of GRBs up to redshift $z$, is given by
\begin{equation}
N(z) = \int_{0}^{z} \frac{dN_{\rm GRB}}{dz'}dz'.
\label{cum_numb}
\end{equation}
So far we have discussed the physical meaning of all terms in  eq. (\ref{psigrb}) except the SFH. Since our aim is to build a model which is independent from the specific form of $\rho_{*}(z)$, we avoid making hypothesis about this quantity. Instead, our intention is to derive $\rho_{*}(z)$ by using PCA and a (mock or real) data set whose data points are $\{z_i,N(z_i)\}$ pairs.


\section{Principal component analysis}
\label{sec:PCA}

The main goal of PCA is the dimensionality reduction of an initial parameter space through the analysis of its internal correlations. Suppose we have a model composed by $P$ parameters. If two of them are highly correlated, they are actually providing the same information. This means that it is possible to rewrite the data in a new parameter space consisting of $P-1$ terms, with minimum loss of information (for a complete review see \cite{jolliffe2002}).
This new set of parameters are recognized as the principal components (PCs), or the eigenvectors of the Fisher information matrix, \textsf{\textbf{F}}.

We postulate that the data set is composed by $N$  independent observations, each one characterized by a Gaussian probability density function,
$f_{i}[g(x_i,\pmb{\beta});G_i,\sigma_{i}]$. In our notation, $x_i$ is a measurement of an independent variable; $G_{i}$ represent the measurements of a quantity $G$ depending on $x_i$; $\sigma_{i}$ is the uncertainty associated with the measures, and
$\pmb{\beta}$ is the parameter vector of the theoretical model. In other words, 
we investigate a specific quantity, $g$, which can be written as a function of the parameters $\beta_i$. In this context, the likelihood function is given by ${\cal L}=\prod_{i=1}^{N}f_i$ and the Fisher matrix is defined as
\begin{equation}
F_{kl}\equiv\left\langle-\frac{\partial^{2}\ln{L(\pmb{\beta})}}
{\partial\beta_{k}\partial\beta_{l}}\right\rangle.\label{eq:FMdefinicao}
\end{equation}
Brackets in eq. (\ref{eq:FMdefinicao}) represent the expectation value.

We can now diagonalize  \textbf{\textsf{F}}, and determine the set of its eigenvectors/PCs, $\pmb{e}$, and eigenvalues, $\pmb{\lambda}$. Following the standard convention, we enumerate $\pmb{e}_i$ from the largest to the smallest associated eigenvalue. 
Our ability to determine the form of each PC is given by $\sigma_{\rm PC_i}=\lambda_i^{-1/2}$.

The set $\pmb{e}$ forms a complete base of uncorrelated vectors. This allows us to use a subspace of $\pmb{e}$, $\pmb{eM}$,  to 
rewrite $g$ as a linear combination of all the elements in $\pmb{eM}$, $g_{rec}(x,\pmb{\alpha})$. The data is then used to find the values of the linear expansion coefficients, $\pmb{\alpha}$ (for a detailed discussion see I2011).

The question of how many PCs should be used in the final reconstruction, or how to choose the dimensionality of $\pmb{eM}$, depends on the particular data set analyzed and our expectation towards them. To provide an idea of how much of the initial information (variance) is included in our plots, we shall order them following their \textit{cumulative percentage of total variance}. A reconstruction with the first $M$ PCs encloses a percentage of this value 
\begin{equation}
t_M=100 \frac{\sum_{i=1}^M \lambda_i}{\sum_{j=1}^{P}\lambda_j}.\label{eq:tM}
\end{equation}
It is important to emphasize that each added  PC brings its associated uncertainty  ($\sigma_{PC_i}$) into the reconstruction. So, although the best-fitted reconstruction converges to the ``real" function as M increases\footnote{This limit holds in ideal situations, where the number of data points largely exceeds the number of initial parameters. If the data set is small or approximately the same size of the initial parameter space, one should also care for overfitting problems.}, the uncertainty associated also raises. As a consequence, the question of how many PCs turns into a matter of what percentage of total variance we are willing to enclose.

\subsection{Star formation history from GRB distribution}

To specify our method of SFH reconstruction from GRB data, let us consider a data set formed by $T$ measurements of the cumulative number of GRB up to redshift $z_i$, $N_{data_i}=N(z_i)$, and its corresponding uncertainty ($\sigma_i$). The likelihood is given by
\begin{eqnarray}
{\cal L}(\pmb{\beta})\propto  \prod_{i=1}^T\exp\left[-\frac{1}{2}\left(\frac{N_{data_i}-N(z_i,\pmb{\beta})}{\sigma_i}\right)^2\right].
\end{eqnarray}
Using equation (\ref{eq:FMdefinicao}), the Fisher matrix components are 
\begin{equation}
F_{k,l}=\sum_{i=1}^T\frac{1}{\sigma_i^2}\frac{\partial N(z_i,\pmb{\beta})}{\partial \pmb{\beta_k}}\frac{\partial N(z_i,\pmb{\beta})}{\partial \pmb{\beta_l}}.
\label{fisher_N}
\end{equation}
The Fisher matrix determination is  now a matter of calculating the derivatives of $N(z,\pmb{\beta})$, for which we use the theoretical prescriptions of Sec. \ref{sec_GRB_rate}. 
Aiming at model independence and simplicity, we model the SFH as 
\begin{equation}
\rho_{*}(z,\pmb{\beta}) = \sum_{i=1}^{n_{\rm bin}}{\beta_i} c_i(z),
\label{rhostep}
\end{equation}
where $\beta_i$ are constants, $n_{\rm bin}$ is the total number of redshift bins, and $c_i(z)$ is a window function which returns 1 if $z_i < z \leq z_{i+1}$ and 0 otherwise. Using this description, we may write any functional form with resolution limited by our computational power.

The derivatives of $N(z)$ can be computed analytically: 
\begin{eqnarray}
\frac{\partial N(z,\pmb{\beta})}{\partial \pmb{\beta_k}}&=&H(J(z)+1-k)
\left[\sum_{i=1}^{J(z)} \delta_{k,i}A(zl_i,zl_{i+1})+\right.\nonumber\\
&&\left.+ \delta_{k,J(z)+1}A(zl_{J(z)+1},z)\right],
\label{eq:Nderiv}
\end{eqnarray}
where $H(x)$ is a step function which returns 0 if $x < 0$ and 1 otherwise, $J(z)$ corresponds to the number of integer bins up to redshift $z$, $\delta_{i,j}$ is the Kronecker delta function, $zl_i$ is the lower bound of the $i$-th redshift bin, and
\begin{eqnarray}
A(z_1,z_2) &\equiv& \frac{\Omega_{obs}\Delta t}{4\mathcal{\pi}}f_{b} f_{GRB}P_{z}\times\nonumber\\
&\times& \int_{z_1}^{z_2}\frac{1}{(1+x)}\frac{dV}{dz}(x)\int_{\log{L_{lim}(x)}}^{\infty}\Phi(L)d\log{L}dx.\nonumber\\
\end{eqnarray}
From these relations the Fisher matrix can be computed and the functional form of the SFH reconstructed through PCA. 
\begin{figure}
\centering
\includegraphics[width=1\columnwidth]{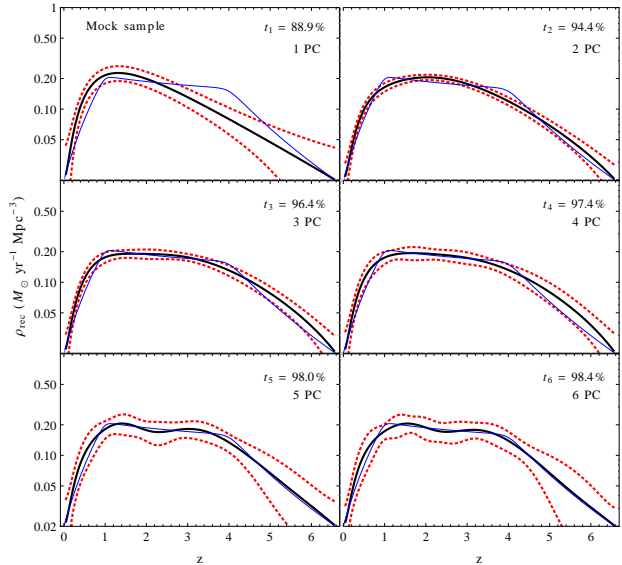}
\caption{PCA reconstructions of SFH obtained from our mock data, using 1 (top-left) to 6 (bottom-right) PCs. The blue-thin line corresponds to our fiducial model $\rho_{\rm fid}$. The black-thick line is the final reconstruction for each case and the red-dashed-thick lines corresponds to $2\sigma$ confidence levels. The inset shows the cumulative percentage of total variance, $t_M$.}
\label{fig:rec_sim_err10_ALL}
\end{figure}

\section{Reconstruction}
\label{sec:reconstruction}
\paragraph*{Mock data}
\label{sec:mock} 

The mock sample is composed of data pairs $\{z_i, N(z_i)\}$, distributed in redshift bins of width $\Delta z=0.1$, where $z_{i}$ represents the middle of each bin. The fiducial model used for the SFH is a simple double-exponential function, $\rho_{\rm fid}$, fitted to numerical results by \citet{li2008},
\begin{equation}
\log{\rho_{\rm fid}(z)} = a+b\log{(1+z)},
\label{rho_star}
\end{equation}
with $(a,b) = (-1.70, 3.30)$ for $z <$ 0.993; $(-0.727, 0.0549)$ for $0.993 < z < 3.80$ and $(2.35,-4.46)$ for $z > 3.80$. We generated 500 simulations, each realization with uncertainty in the determination of $N(z)$ set to unit and containing 65 redshift bins. After we generated the mock sample, the information about $\rho_{\rm fid}(z)$ is discarded.  Our goal from now on is to re-obtain the functional form of eq. (\ref{rho_star}) from PCA.

The Fisher matrix is obtained as described above, assuming an observing time $\Delta t_{\rm MS}=1$ yr. As the mock sample purpose is to test the procedure under an ideal scenario, we did not include uncertainties in the parameters of eq. (\ref{psigrb}). Having obtained the PCs, we can rewrite the SFH as
\begin{equation}
\rho_{rec}(z)=\rho_c+\sum_{i=1}^M\alpha_i\pmb{e_i}(z),
\end{equation}
where $\alpha_i$ and $\rho_c$ are constants to be determined and $M$ is the number of PCs we choose to use in the reconstruction. The simulated data points are then used to find the appropriate values for the parameters $\alpha_i$ and $\rho_c$ as those that minimize the expression
\begin{equation}
\chi^2(\pmb{\alpha})\propto\sum_{i=1}^{T_{\rm MS}}\frac{\left(N_{i;{\rm data}}-N_{\rm rec}(z_i;\rho_c,\pmb{\alpha})\right)^2}{2\sigma_i^2},
\end{equation}
where $\sigma_i =1$ for all redshift bins. 
The reconstructions obtained using 1 to 6 PCs are shown in Fig. \ref{fig:rec_sim_err10_ALL}. The uncertainty in the final reconstruction was calculated by a quadrature sum that includes the parameters $\sigma_{\rm PC_i}$ and the uncertainty in the determination of parameters $\pmb{\alpha} (\sigma_{\alpha_i})$ and $\rho_c (\sigma_{\rho_c})$.

From Fig. \ref{fig:rec_sim_err10_ALL} we can appreciate the success of the procedure in reconstructing the underlying unknown SFH in an ideal scenario, with increasing agreement as the number of PCs raises. Confidence levels also become wider as $M$ increases, with the only exception of the reconstruction with 1 PC. Since $\sigma_{\rho_c}$ dominates the errors due to the limited freedom to fit the second peak of the fiducial model with only 1 PC. With 2 PCs fitting the second peak becomes easier, and as a consequence,  the magnitude of $\sigma_{\rho_c}$ decreases to levels below those of $\sigma_{\rm PC_i}$. 


\paragraph*{\textit{Swift} data}
\label{sec:SWIFT}
After validating PCA reconstruction under ideal conditions, we turn to the use of currently available \textit{Swift} data, and compare these results with independent measurements of SFH from the literature. First, we need to properly choose our data set. Since only GRBs with measured redshifts can be used in our analysis, the question of how the redshift measurements were obtained must be examined carefully. 

\begin{figure}
\includegraphics[width=1\columnwidth]{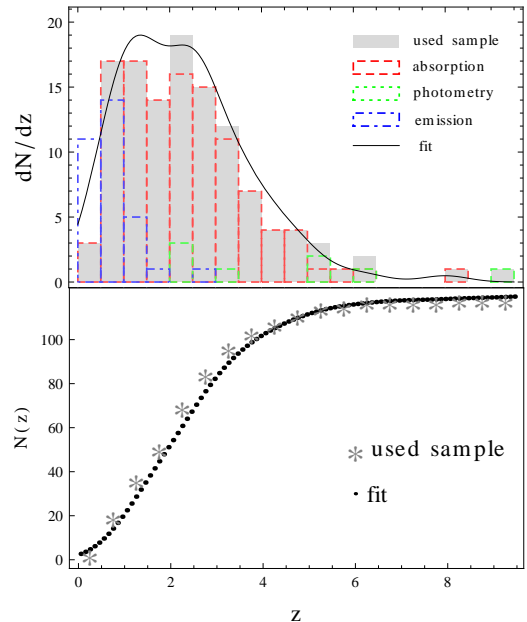}
\caption{\textbf{Top}: Histogram showing the measured redshift distribution of 120 GRBs detected by {\it Swift} from 2005 to 2010, divided by redshift measurement methods: absorption (red-dashed), photometry (green-dotted) and emission (blue-dot-dashed).  The gray region corresponds to the redshift distribution of data we used (absorption + photometry). The solid black line shows the fit to the used data distribution. 
\textbf{Bottom}: The cumulative distributions constructed from {\it Swift} data (gray stars) and from our fitted distribution function (black points).}
\label{fig:SWIFT_hist}
\end{figure}

GRBs redshifts are generally obtained from optical afterglow spectra using absorption lines or photometry, or from the spectrum of the host galaxy using emission lines. As pointed by \citet{wanderman2010}, different methods yield different redshift distributions: a visual inspection of Fig. \ref{fig:SWIFT_hist} illustrates this point. Most noticeably, the GRB redshift distribution determined from their hosts lacks very high-$z$ events.   Moreover, emission (and to a lesser extent, absorption) lines are susceptible to a selection effect known as the ``redshift desert" in the range $1.1 < z < 2.1$ \citep{Fiore2007,Coward2009}. Additional bias sources are preliminary discussed by \citet{Malesani2009}. To avoid systematic errors affecting the overall redshift distribution, our data sample is composed by 120 \textit{Swift} GRBs  with redshift determined from absorption lines and photometry (gray region in Fig. \ref{fig:SWIFT_hist}, top panel).

The next step is to choose the appropriate redshift bin width. In principle, the quality of the reconstruction should increase with the number of bins. However, as GRB are discrete events, if we pick a bin width based on the available data (for example, in such a way that each bin has at least one GRB), the bins will be too wide ($\approx 1$). In this case, the assumption that the SFH is constant inside the bin will not hold, leading to reconstructions with bad resolution. To overcome this limitation we performed a gaussian  kernel fit\footnote{A non-parametric estimate of the PDF obtained from a linearly interpolated version of $\frac{1}{n h}\sum_{i=1}^nk\left(\frac{x-x_i}{h}\right)$ for a kernel $k(x)$, bin width $h$ and a total of $n$ bins.} to the data (black line in Fig. \ref{fig:SWIFT_hist}, top panel). Now we have a continuous probability distribution function (PDF) for $dN/dz$, which follows the real data distribution and allows us to set the bin width as small as required. We kept $\Delta z=0.1$ and use the PDF to calculate the cumulative number of observed GRBs in each bin. The comparison between the real data cumulative distribution and the one calculated via the fitted PDF are shown in Fig. \ref{fig:SWIFT_hist}, bottom panel.

\begin{figure}
\centering
\includegraphics[width=1\columnwidth]{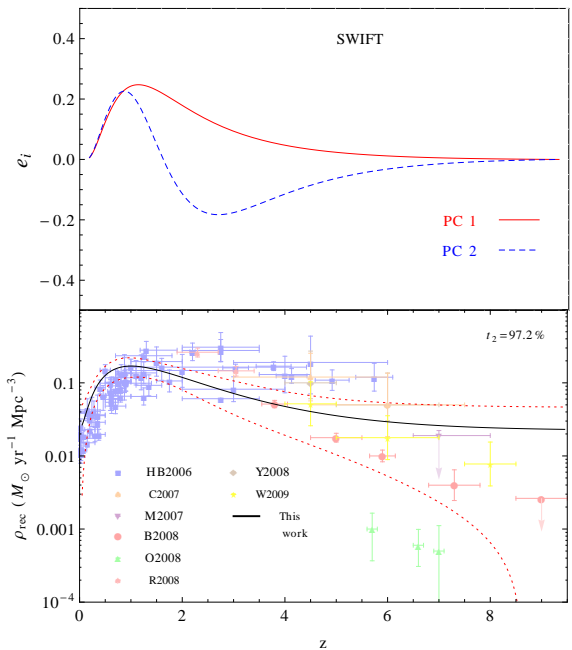}
\caption{\textbf{Top}: First (red solid) and second (blue dashed) PCs from \textit{Swift}. \textbf{Bottom}: PCA reconstruction from \textit{Swift} data using 2 PCs, compared with independent SFH determinations (light points, not used in our calculations). The black solid line is the PCA best-fit reconstruction using 2 PCs; the red dashed lines correspond to $2\sigma$ confidence levels. The inset shows the cumulative percentage of total variance, $t_M$.}
\label{fig:rec_data_err10_ALL}
\end{figure}

The Fisher matrix is calculated using $\Delta t_{\rm SW}=6$ yr of observation time. The parameters $\sigma_i$ were obtained by summing in quadrature the uncertainties in the quantities involved in eq. (\ref{psigrb}). Fig. \ref{fig:rec_data_err10_ALL} shows the first 2 PCs and the corresponding reconstruction using both of them, which already encloses more than $97\%$ of total variance. In the lower panel, the points correspond to completely independent measurements from the literature. 
These data points are shown only for comparison purposes and have not been used in our calculations.

\section{Discussion}
\label{sec_disc}

We have proposed the use of PCA as a powerful tool to reconstruct the cosmic star formation history exploiting the measured gamma-ray burst redshift distribution. The procedure was successfully validated using synthetic data and next applied to actual {\it Swift} data (Fig. 3).

It is important to highlight that the approach is completely independent of the initial choice of the theoretical model parameter vector, $\pmb{\beta}$. This has the obvious advantage of avoiding any a priori hypothesis on the SFH, $\rho_*(z)$. However, the degeneracy between $\rho_*(z)$ and any other factor we are failing to take into account cannot be removed, i.e. the reconstructed SFH contains also the behavior of all the agents influencing the determination of $N(z)$ and not included in the model.

For example,  \citet{langer2006,woosley2006} have argued that GRB progenitors will
have a low metallicity. Such an effect would be a consequence of the mass and angular momentum loss  caused by winds in high-metallicity stars. This would prevent such stars of becoming GRBs and consequently,  change their expected redshift distribution  \citep{salvaterra2007a,salvaterra2007b,li2008}. 
We implicitly considered that this and other such effects will span within  the error bars in our analysis.

In spite of the remaining uncertainties, which are probably less severe than those affecting other methods aimed at recovering the high-$z$ tail of the SFH, there are robust indications that we can gather from the analysis of our results. The first is that the combination of GRB data and PCA suggest that the level of star formation activity at $z\approx 9.4$ could have been  already as high as the present-day one ($\approx 0.01 M_\odot$ yr$^{-1}$ Mpc$^{-3}$). This is a factor 3-5 times higher than deduced from high-$z$ galaxy searches through drop-out techniques, similarly to the trend found by \citet{Yonetoku2004}. If true, this might alleviate the long-standing problem of a photon-starving reionization; it might also indicate that galaxies accounting for most of the star formation activity at high redshift go undetected by even the most deep searches. Finally it is worth noticing that a sustained high-$z$ star formation activity is consistent with predictions of reionization models that simultaneously match a number of observable experimental constraints as the Gunn-Peterson effect, Thomson free-electron optical depth, Lyman Limit Systems statistics etc. (\citet{Choudhury2006}, \citet{Bolton2007}). Given the expected large growth of GRB detections from the next generation of instruments, the method proposed here promises to become one of the most suitable and reliable tools to constrain the star formation activity in the young Universe.

\section*{Acknowledgements}
\begin{footnotesize}
We thank K. Ioka, R. Salvaterra and N. Yoshida for useful comments. E.E.O.I. thanks  CAPES (1313-10-0) for financial support. R.S.S.  thanks  CNPq (200297/2010-4) for financial support.  AF acknowledges support from IPMU where this research started. This work was supported by WPI Initiative, MEXT, Japan.
\end{footnotesize}

\bibliographystyle{mn2e}

\begin{thebibliography}{}

\bibitem[\protect\citeauthoryear{{Bolton} \& {Haehnelt}}{{Bolton} \&
  {Haehnelt}}{2007}]{Bolton2007}
{Bolton} J.~S.,  {Haehnelt} M.~G.,  2007, \mnras, 381, L35

\bibitem[\protect\citeauthoryear{{Bouwens},  \& {et al.}}{{Bouwens}
  et~al.}{2008}]{bouwens2008}
{Bouwens} R.~J.,     {et al.} 2008, \apj, 686, 230

\bibitem[\protect\citeauthoryear{{Bromm} \& {Loeb}}{{Bromm} \&
  {Loeb}}{2002}]{bromm2002}
{Bromm} V.,  {Loeb} A.,  2002, \apj, 575, 111

\bibitem[\protect\citeauthoryear{{Bromm} \& {Loeb}}{{Bromm} \&
  {Loeb}}{2006}]{bromm2006}
{Bromm} V.,  {Loeb} A.,  2006, \apj, 642, 382

\bibitem[\protect\citeauthoryear{{Campisi}, {Li} \& {Jakobsson}}{{Campisi}
  et~al.}{2010}]{campisi2010}
{Campisi} M.~A.,  {Li} L.-X.,    {Jakobsson} P.,  2010, \mnras, 407, 1972

\bibitem[\protect\citeauthoryear{{Campisi}, {Maio}, {Salvaterra} \&
  {Ciardi}}{{Campisi} et~al.}{2011}]{campisi2011}
{Campisi} M.~A.,  {Maio} U.,  {Salvaterra} R.,    {Ciardi} B.,  2011, ArXiv
  e-prints

\bibitem[\protect\citeauthoryear{{Chary}, {Berger} \& {Cowie}}{{Chary}
  et~al.}{2007}]{chary2007}
{Chary} R.,  {Berger} E.,    {Cowie} L.,  2007, \apj, 671, 272

\bibitem[\protect\citeauthoryear{{Choudhury} \& {Ferrara}}{{Choudhury} \&
  {Ferrara}}{2006}]{Choudhury2006}
{Choudhury} T.~R.,  {Ferrara} A.,  2006, \mnras, 371, L55

\bibitem[\protect\citeauthoryear{{Ciardi} \& {Loeb}}{{Ciardi} \&
  {Loeb}}{2000}]{ciardi2000}
{Ciardi} B.,  {Loeb} A.,  2000, \apj, 540, 687

\bibitem[\protect\citeauthoryear{{Conselice}, {Vreeswijk}, {Fruchter}, {Levan},
  {Kouveliotou}, {Fynbo}, {Gorosabel}, {Tanvir} \& {Thorsett}}{{Conselice}
  et~al.}{2005}]{conselice2005}
{Conselice} C.~J.,  {Vreeswijk} P.~M.,  {Fruchter} A.~S.,  {Levan} A.,
  {Kouveliotou} C.,  {Fynbo} J.~P.~U.,  {Gorosabel} J.,  {Tanvir} N.~R.,
  {Thorsett} S.~E.,  2005, \apj, 633, 29

\bibitem[\protect\citeauthoryear{{Coward}}{{Coward}}{2009}]{Coward2009}
{Coward} D.~M.,  2009, \mnras, 393, L65

\bibitem[\protect\citeauthoryear{{Cucchiara},  \& {et al.}}{{Cucchiara}
  et~al.}{2011}]{Cucchiara2011}
{Cucchiara} A.,     {et al.} 2011, arXiv:astro-ph/1105.4915

\bibitem[\protect\citeauthoryear{{de Souza}, {Yoshida} \& {Ioka}}{{de Souza}
  et~al.}{2011}]{rafael2011}
{de Souza} R.~S.,  {Yoshida} N.,    {Ioka} K.,  2011, arXiv:astro-ph/1105.2395

\bibitem[\protect\citeauthoryear{{Fiore},  \& {et al.}}{{Fiore}
  et~al.}{2007}]{Fiore2007}
{Fiore} F.,     {et al.} 2007, \aap, 470, 515

\bibitem[\protect\citeauthoryear{{Fynbo},  \& {et al.}}{{Fynbo}
  et~al.}{2006}]{fynbo2006}
{Fynbo} J.~P.~U.,     {et al.} 2006, \aap, 451, L47

\bibitem[\protect\citeauthoryear{{Gal-Yam},  \& {et al.}}{{Gal-Yam}
  et~al.}{2006}]{gal2006}
{Gal-Yam} A.,     {et al.} 2006, \apj, 639, 331

\bibitem[\protect\citeauthoryear{{Guetta}, {Piran} \& {Waxman}}{{Guetta}
  et~al.}{2005}]{guetta2005}
{Guetta} D.,  {Piran} T.,    {Waxman} E.,  2005, \apj, 619, 412

\bibitem[\protect\citeauthoryear{{Hopkins} \& {Beacom}}{{Hopkins} \&
  {Beacom}}{2006}]{hopkins2006}
{Hopkins} A.~M.,  {Beacom} J.~F.,  2006, \apj, 651, 142

\bibitem[\protect\citeauthoryear{{Huterer} \& {Starkman}}{{Huterer} \&
  {Starkman}}{2003}]{huterer2003}
{Huterer} D.,  {Starkman} G.,  2003, Physical Review Letters, 90, 031301

\bibitem[\protect\citeauthoryear{{Ishida} \& {de Souza}}{{Ishida} \& {de
  Souza}}{2011}]{ishida2011}
{Ishida} E.~E.~O.,  {de Souza} R.~S.,  2011, \aap, 527, A49+

\bibitem[\protect\citeauthoryear{{Jarosik},  \& {et al.}}{{Jarosik}
  et~al.}{2011}]{jarosik2010}
{Jarosik} N.,     {et al.} 2011, \apjs, 192, 14

\bibitem[\protect\citeauthoryear{{Jolliffe}}{{Jolliffe}}{2002}]{jolliffe2002}
{Jolliffe} I.~T.,  2002, {Principal component analysis}

\bibitem[\protect\citeauthoryear{{Kocevski}, {West} \& {Modjaz}}{{Kocevski}
  et~al.}{2009}]{kocevski2009}
{Kocevski} D.,  {West} A.~A.,    {Modjaz} M.,  2009, \apj, 702, 377

\bibitem[\protect\citeauthoryear{{Lamb} \& {Reichart}}{{Lamb} \&
  {Reichart}}{2000}]{lamb2000}
{Lamb} D.~Q.,  {Reichart} D.~E.,  2000, \apj, 536, 1

\bibitem[\protect\citeauthoryear{{Langer} \& {Norman}}{{Langer} \&
  {Norman}}{2006}]{langer2006}
{Langer} N.,  {Norman} C.~A.,  2006, \apjl, 638, L63

\bibitem[\protect\citeauthoryear{{Lehnert},  \& {et al.}}{{Lehnert}
  et~al.}{2010}]{Lehnert2010}
{Lehnert} M.~D.,     {et al.} 2010, \nat, 467, 940

\bibitem[\protect\citeauthoryear{{Li}}{{Li}}{2008}]{li2008}
{Li} L.,  2008, \mnras, 388, 1487

\bibitem[\protect\citeauthoryear{{Malesani},  \& {et al.}}{{Malesani}
  et~al.}{2009}]{Malesani2009}
{Malesani} D.,     {et al.} 2009, \apjl, 692, L84

\bibitem[\protect\citeauthoryear{{Mannucci},  \& {et al.}}{{Mannucci}
  et~al.}{2007}]{mannucci2007}
{Mannucci} F.,     {et al.} 2007, \aap, 461, 423

\bibitem[\protect\citeauthoryear{{Maturi} \& {Mignone}}{{Maturi} \&
  {Mignone}}{2009}]{maturi2009}
{Maturi} M.,  {Mignone} C.,  2009, \aap, 508, 45

\bibitem[\protect\citeauthoryear{{Mitra}, {Choudhury} \& {Ferrara}}{{Mitra}
  et~al.}{2010}]{mitra2010}
{Mitra} S.,  {Choudhury} T.~R.,    {Ferrara} A.,  2010,
  arXiv:astro-ph/1011.2213

\bibitem[\protect\citeauthoryear{{Ota},  \& {et al.}}{{Ota}
  et~al.}{2008}]{ota2008}
{Ota} K.,     {et al.} 2008, \apj, 677, 12

\bibitem[\protect\citeauthoryear{{Porciani} \& {Madau}}{{Porciani} \&
  {Madau}}{2001}]{porciani2001}
{Porciani} C.,  {Madau} P.,  2001, \apj, 548, 522

\bibitem[\protect\citeauthoryear{{Price},  \& {et al.}}{{Price}
  et~al.}{2006}]{price2006}
{Price} P.~A.,     {et al.} 2006, \apj, 645, 851

\bibitem[\protect\citeauthoryear{{Prochaska},  \& {et al.}}{{Prochaska}
  et~al.}{2007}]{prochaska2007}
{Prochaska} J.~X.,     {et al.} 2007, \apj, 666, 267

\bibitem[\protect\citeauthoryear{{Reddy},  \& {et al.}}{{Reddy}
  et~al.}{2008}]{reddy2008}
{Reddy} N.~A.,     {et al.} 2008, \apjs, 175, 48

\bibitem[\protect\citeauthoryear{{Salvaterra},  \& {et al.}}{{Salvaterra}
  et~al.}{2008}]{Salvaterra2008}
{Salvaterra} R.,     {et al.} 2008, \mnras, 385, 189

\bibitem[\protect\citeauthoryear{{Salvaterra}, {Campana}, {Chincarini},
  {Tagliaferri} \& {Covino}}{{Salvaterra} et~al.}{2007}]{salvaterra2007b}
{Salvaterra} R.,  {Campana} S.,  {Chincarini} G.,  {Tagliaferri} G.,
  {Covino} S.,  2007, \mnras, 380, L45

\bibitem[\protect\citeauthoryear{{Salvaterra} \& {Chincarini}}{{Salvaterra} \&
  {Chincarini}}{2007}]{salvaterra2007a}
{Salvaterra} R.,  {Chincarini} G.,  2007, \apjl, 656, L49

\bibitem[\protect\citeauthoryear{{Salvaterra}, {Della Valle}, {Campana} \& {et
  al.}}{{Salvaterra} et~al.}{2009}]{Salvaterra2009}
{Salvaterra} R.,  {Della Valle} M.,  {Campana} S.,    {et al.} 2009, \nat, 461,
  1258

\bibitem[\protect\citeauthoryear{{Savaglio}}{{Savaglio}}{2006}]{savaglio2006}
{Savaglio} S.,  2006, New Journal of Physics, 8, 195

\bibitem[\protect\citeauthoryear{{Schneider},  \& {et al.}}{{Schneider}
  et~al.}{2006}]{schneider2006}
{Schneider} R.,     {et al.} 2006, \mnras, 369, 1437

\bibitem[\protect\citeauthoryear{{Tanvir},  \& {et al.}}{{Tanvir}
  et~al.}{2009}]{Tanvir2009}
{Tanvir} N.~R.,     {et al.} 2009, \nat, 461, 1254

\bibitem[\protect\citeauthoryear{{Totani},  \& {et al.}}{{Totani}
  et~al.}{2006}]{totani2006}
{Totani} T.,     {et al.} 2006, \pasj, 58, 485

\bibitem[\protect\citeauthoryear{{Totani}}{{Totani}}{1997}]{totani1997}
{Totani} T.,  1997, \apjl, 486, L71+

\bibitem[\protect\citeauthoryear{{Wanderman} \& {Piran}}{{Wanderman} \&
  {Piran}}{2010}]{wanderman2010}
{Wanderman} D.,  {Piran} T.,  2010, \mnras, 406, 1944

\bibitem[\protect\citeauthoryear{{Wang} \& {Dai}}{{Wang} \&
  {Dai}}{2009}]{wang2009}
{Wang} F.~Y.,  {Dai} Z.~G.,  2009, \mnras, 400, L10

\bibitem[\protect\citeauthoryear{Woosley \& Heger}{Woosley \&
  Heger}{2006}]{woosley2006}
Woosley S.~E.,  Heger A.,  2006, The Astrophysical Journal, 637, 914

\bibitem[\protect\citeauthoryear{{Yonetoku},  \& {et al.}}{{Yonetoku}
  et~al.}{2004}]{Yonetoku2004}
{Yonetoku} D.,     {et al.} 2004, \apj, 609, 935

\bibitem[\protect\citeauthoryear{{Y{\"u}ksel},  \& {et al.}}{{Y{\"u}ksel}
  et~al.}{2008}]{yuksel2008}
{Y{\"u}ksel} H.,     {et al.} 2008, \apjl, 683, L5

\end{thebibliography}

\bsp

\label{lastpage}
\end{document}